\documentclass[
journal=jacsat, 
manuscript=article]{achemso}

\usepackage[version=3]{mhchem} 

\author{Alexey A. Shvets}
\affiliation[MIT]{Institute for Medical Engineering and Science, Massachusetts Institute of Technology; Cambridge, MA 02142, USA}

\author{Maria P. Kochugaeva}
\affiliation[Yale]
{Department of Biomedical Engineering and System Biology Institute Yale University West Haven, CT, 06516, USA}

\author{Anatoly B. Kolomeisky}
\email{tolya@rice.edu}
\affiliation[Rice]
{Department of Chemistry, Department of Chemical and Biomolecular Engineering, and Center for Theoretical Biological Physics, Rice University, Houston, Texas 77005, USA}

\title[\texttt{achemso} demonstration]
{Mechanisms of Protein Search for Targets on DNA: Theoretical Insights}

\begin{document}

\begin{abstract}
Protein-DNA interactions are critical for the successful functioning of all natural systems. The key role in these interactions is played by processes of protein search for specific sites on DNA. Although it has been studied for many years, only recently microscopic aspects of these processes became more clear. In this work, we present a review on current theoretical understanding of the molecular mechanisms of the protein target search. A comprehensive discrete-state stochastic method to explain the dynamics of the protein search phenomena is introduced and explained.  Our theoretical approach utilizes a first-passage analysis and it takes into account the most relevant physical-chemical processes. It is able to describe many fascinating features of the protein search, including unusually high effective association rates, high selectivity and specificity, and the robustness in the presence of crowders and sequence heterogeneity.
\end{abstract}

\section{Introduction}

Dynamical nature of underlying processes is what distinguishes the living systems from other processes. \cite{alberts,lodish}. Biological processes constantly involve time-dependent fluxes of energy and materials, which makes them strongly deviating from  equilibrium as long as organisms are alive. This implies that the concepts of equilibrium thermodynamics have limited applications for biological systems, while the role of methods that study the dynamical transformations is much more important \cite{phillips}. In this review, we present our theoretical views on dynamic aspects of the protein-DNA interactions, which dominate in biological systems. Our approach is based on explicit calculations  of dynamic properties via a first-passage probabilities analysis. The first-passage ideas have been already widely utilized in studies of various complex processes in Chemistry, Physics and Biology \cite{vankampen,redner}. We employ these ideas in developing a discrete-state stochastic framework for analyzing the dynamics of protein search for specific targets on DNA.

It is known that the beginning of most biological processes is associated with specific protein molecules binding to specific target sequences on DNA because these events initiate the cascades of corresponding biochemical and biophysical processes \cite{alberts,lodish,phillips}. For example, to activate or to repress a gene the corresponding transcription factor proteins must bind first to the gene promoter's region  \cite{alberts,lodish}. This fundamental aspect of protein-DNA interactions has been studied extensively by various experimental and theoretical methods \cite{riggs70,berg81,berg85,gowers05,halford04,mirny09,kolomeisky11,hu06,hu08,bauer12,sheinman12,veksler13,cuculis15,kochugaeva16,kochugaeva17,kochugaeva17a,slutsky04,tafvizi11p53,shvets15,shvets16,shvets15b,shvets16b,zandarashvili15,esadze14,hammar12,mahmutovic15,reingruber11,shin18,lange15a,lange15b,shvets17,benichou09,koslover11}.  A special attention was devoted to understanding the dynamics of the protein search for specific targets on DNA. Many ideas have been proposed and critically discussed, but only recently a clear molecular picture of the underlying processes started to emerge \cite{mirny09,kolomeisky11,veksler13}.

Large amount of experimental observations on protein search phenomena, which mostly come from the single-molecule measurements, suggests that it is a complex dynamic phenomenon which combines three-dimensional (in the bulk solution) and one-dimensional (on the DNA chain) motions \cite{gowers05, halford04,mirny09,kolomeisky11,sheinman12}.  But the most paradoxical observation is that, although the protein molecules spend most of the search time ($\ge$90-99$\%$) on the DNA chain where they diffuse very slowly, they still can find the targets very fast, in some cases much faster than the bulk diffusion would allow \cite{halford04, mirny09, kolomeisky11}. For example, the measured association rate for {\it lac}-repressor was $\sim10^{10} M^{-1} s^{-1}$  (two orders of magnitude faster than the diffusion limit!) \cite{riggs70}, and many other experimentally determined protein-DNA association rates were also astonishingly high in comparison to typical biological binding rates.  This is known as a {\it facilitated diffusion}. Several theoretical ideas on the origin of the facilitated diffusion, including lowering of dimensionality, electrostatic effects, correlations between 3D and 1D motions, conformational transitions, bending fluctuations, and hydrodynamics effects have been explored  and discussed \cite{halford04,mirny09,kolomeisky11}. However, theoretical analysis shows that none of these mechanisms can fully explain the facilitated diffusion in the protein search \cite{veksler13}. To understand the dynamic aspects  of protein-DNA interactions, we developed a discrete-state stochastic framework to take into account the most relevant physical-chemical processes in the system. The application of the first-passage probabilities method allows us also to explicitly evaluate the dynamic properties and to clarify dynamic aspects of the protein-DNA interactions.

It is important to note that although there are still different opinions on the theoretical foundations of the protein search phenomena, in this work we mostly present our views on these problems, which, of course, are subjective. In addition,  there are many theoretical advances in our understanding of the protein search dynamics, but we will concentrate only on few of them in order to explain better the underlying molecular processes. Furthermore, there is a huge number of investigations on the protein target search phenomena. Our goal is not to cover all studies and all existing views but to present a clear theoretical picture of these processes as we understand it now.

\section{Simplest Discrete-State Stochastic Model of the Protein Target Search}

\begin{figure}
\centerline{\includegraphics[width=8.8cm]{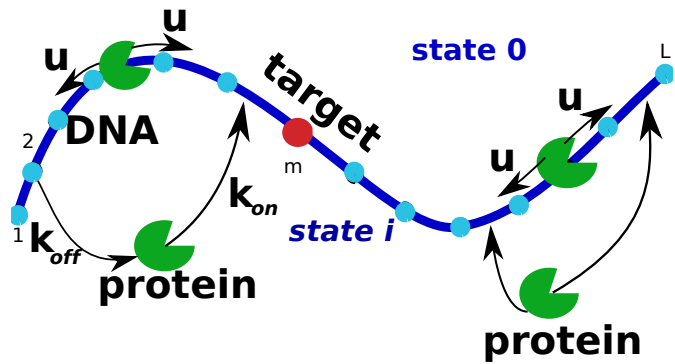}}
\caption{A schematic view of a minimal discrete-state stochastic  model of the protein search for targets om DNA. The DNA chain has $L-1$ non-specific binding sites and one specific target site. A protein molecule can diffuse along the DNA segment with a rate $u$ in both directions. It can also associate to  DNA  from the bulk solution (labeled as state $0$) with a rate $k_{on}$ or it can dissociate back to the solution with a rate  $k_{off}$. The search is finished when the protein binds to the target site at the position $m$ for the first time.}
\label{Fig1}
\end{figure}

Experiments clearly indicate that during the search the protein molecule is alternating between freely diffusing behavior in the solution around the DNA chain and non-specific associations to DNA, which also include scanning the DNA chain \cite{halford04,mirny09,kolomeisky11}. The process is completed when the protein molecule reaches the specific target sequence on DNA for the first time. Stimulated by this observations, we start with a simplest minimal model of the protein search as presented in Figure \ref{Fig1}. It is important to note that, in contrast to other theoretical approaches \cite{halford04,mirny09,benichou09,bauer12}, this method is based on a discrete-state stochastic description of the system. This is a more realistic view of early stages of protein-DNA interactions because of intrinsically discrete nature of molecular interactions in these systems.

In this simple model, we consider a single protein molecule and a single DNA molecule with a single target site: see Figure \ref{Fig1}. The DNA chain is viewed as having $L$ discrete binding sites, and one of them at the position $m$ is considered to be the target for the protein molecule. Because the diffusion of the proteins in the bulk is usually fast, all solutions states for the protein are combined into one state that we label as a state $0$ (Figure \ref{Fig1}). It is assumed that from the bulk solution the protein molecule can bind with equal probability to any site on DNA, and the total association rate to DNA is equal to $k_{on}$, while the dissociation rate from DNA is $k_{off}$. The non-specifically bound proteins can diffuse without bias along the DNA contour in any direction with a rate $u$ (see Figure \ref{Fig1}). Since the search process ends as soon as the protein molecule arrives to the specific site for the first time, we introduce a function $F_{n}(t)$, which is  defined as a probability density function of reaching the site $m$ (the target site) for the first time  at time $t$ if at $t=0$ the protein started in the state $n$ ($n=0$ is the bulk solution, and $n=1,...,L$ are the protein-DNA bound states). This function is also known as a first-passage probability density function \cite{vankampen,redner}. To compute these first-passage probabilities, we utilize backward master equations that describe the temporal evolution of these quantities \cite{vankampen,redner,veksler13},
\begin{equation}\label{eq1}
\frac{dF_{n}(t)}{d t}=u\left[ F_{n+1}(t)+F_{n-1}(t)\right] +k_{off} F_{0}(t) -(2u+k_{off}) F_{n}(t),
\end{equation}
for $2\le n \le L-1$, while at the boundaries ($n=1$ or $n=L$) we have
\begin{equation}\label{eq2}
\frac{dF_{1}(t)}{d t}=u F_{2}(t) +k_{off} F_{0}(t) -(u+k_{off}) F_{1}(t),
\end{equation}
and
\begin{equation}\label{eq3}
\frac{dF_{L}(t)}{d t}=u F_{L-1}(t) +k_{off} F_{0}(t) -(u+k_{off}) F_{L}(t).
\end{equation}
For the state $n=0$, the backward master equation is different,
\begin{equation}\label{eq4}
\frac{dF_{0}(t)}{d t}=\frac{k_{on}}{L} \sum_{n=1}^{L} F_{n}(t) -k_{on} F_{n}(t).
\end{equation}
Here we used the fact that the rate to bind to any site on DNA is $k_{on}/L$, so that the total association rate is equal to $k_{on}$. In addition, the initial conditions require that $F_{m}(t)=\delta(t)$ and $F_{n \ne m}(t=0)=0$. This means that if the protein molecule starts at the target site $m$ the search is immediately accomplished.

It is important to explain the physical meaning of the backward master equations because they differ from classical forward master equations widely employed in Chemical Kinetics. It can be easily seen that all trajectories that start at the state $n$ and finish at the target site $m$ can be divided into several groups. For example, for $2 \le n \le L-1$  all trajectories starting at $n$ can be divided into three groups: 1) passing via the state $n-1$, 2) passing via the state $n+1$ or 3) passing via the state $0$ in the next time step. The fractions of those trajectories are given by $u/(2u+k_{off})$,  $u/(2u+k_{off})$ and $k_{off}/(2u+k_{off})$, respectively. Equation (\ref{eq1}) describes this partition of the trajectories in the time-dependent manner because the first-passage probability flux to the target is determined by these trajectories. Thus, the backward master equations reflect the temporal evolution of the first-passage probabilities.

The most convenient way to analyze the dynamics in the system is to use Laplace representations of the first-passage probability functions, $\widetilde{F_{n}(s)} \equiv \int_{0}^{\infty} e^{-st} F_{n}(t) dt$. Then Equations (\ref{eq1}),(\ref{eq2}), (\ref{eq3}) and (\ref{eq4}) can be written as simpler algebraic expressions:
\begin{equation}
(s+2u+k_{off}) \widetilde{F_{n}(s)}=u\left[\widetilde{F_{n+1}(s)}+\widetilde{F_{n-1}(s)}\right]+k_{off} \widetilde{F_{0}(s)};
\end{equation}
\begin{equation}
(s+u+k_{off}) \widetilde{F_{1}(s)}=u\widetilde{F_{2}(s)}+k_{off} \widetilde{F_{0}(s)};
\end{equation}
\begin{equation}
(s+u+k_{off}) \widetilde{F_{L}(s)}=u\widetilde{F_{L-1}(s)}+k_{off} \widetilde{F_{0}(s)};
\end{equation}
\begin{equation}
(s+k_{on}) \widetilde{F_{0}(s)}=\frac{k_{on}}{L}\sum_{n=1}^{L}\widetilde{F_{n}(s)}.
\end{equation}
In addition, from the initial conditions we have $\widetilde{F_{m}(s)}=1 $. These equations are solved assuming that the general form of the solution is $\widetilde{F_{n}(s)}=A y^{n} +B$, where the unknown coefficients $A$, $y$ and $B$ are determined from the initial and boundary conditions \cite{veksler13}. One could argue that the target site $m$ divides the DNA molecule into two homogeneous segments ($1\le n \le m$ and $m \le n \le L$), which can be considered separately. It was shown \cite{veksler13} that this approach leads to explicit expressions for the first-passage probability functions. Specifically, one obtains
\begin{equation}
 \widetilde{F_{0}(s)}=\frac{k_{on}(k_{off}+s)S_{1}(s)}{L s(k_{off}+k_{on}+s)+k_{off}k_{on}S_{1}(s)}, 
\end{equation}
with an auxiliary function $S_{1}(s)$ defined as
\begin{equation}
    S_{1}(s)=\frac{y(1+y)(y^{-L}-y^{L})}{(1-y)(y^{1-m}+y^{m})(y^{m-L}+y^{1+L-m})};
\end{equation}
and with the parameters $y$ and $B$ given by
\begin{equation}\label{eq_y}
  y=\frac{s+2u+k_{off}-\sqrt{(s+2u+k_{off})^{2}-4u^{2}}}{2u};
\end{equation}
\begin{equation}
    B=\frac{k_{off}\widetilde{F_{0}(s)}}{(k_{off}+s)}.
\end{equation}

Explicit expressions for the first-passage probabilities provide a full dynamic description of the protein search processes and any relevant quantities can be easily computed. For example, the mean search time from the bulk solution, which is inversely proportional to the chemical association rate for the specific target site,  can be found from\cite{veksler13},
\begin{equation}\label{eq13}
    T_{0} \equiv -\frac{\partial \widetilde{F_{0}(s)}}{\partial s} \bigg\vert_{s=0}=\frac{1}{k_{on}}\frac{L}{S_{1}(0)}+\frac{1}{k_{off}}\frac{L-S_{1}(0)}{S_{1}(0)}.
\end{equation}
This result has a very clear physical meaning. Here the parameter $S_{1}(0)$ describes the average number of distinct sites that the protein molecule scans during each visit to DNA while searching for the single specific site. Then, on average, to find the target the protein must make $L/S_{1}(0)$ visits to DNA because during every association $S_{1}(0)$ DNA sites are checked. Each visit, on average, lasts $1/k_{on}$ while the protein scans for the target diffusing along the DNA chain. The protein also makes $L/S_{1}(0)-1$ dissociations back into the solution. The number of dissociation events is smaller by one than the number of association events because the last binding to DNA leads to finding the specific site.

\begin{figure}
\centerline{\includegraphics[width=8.8cm]{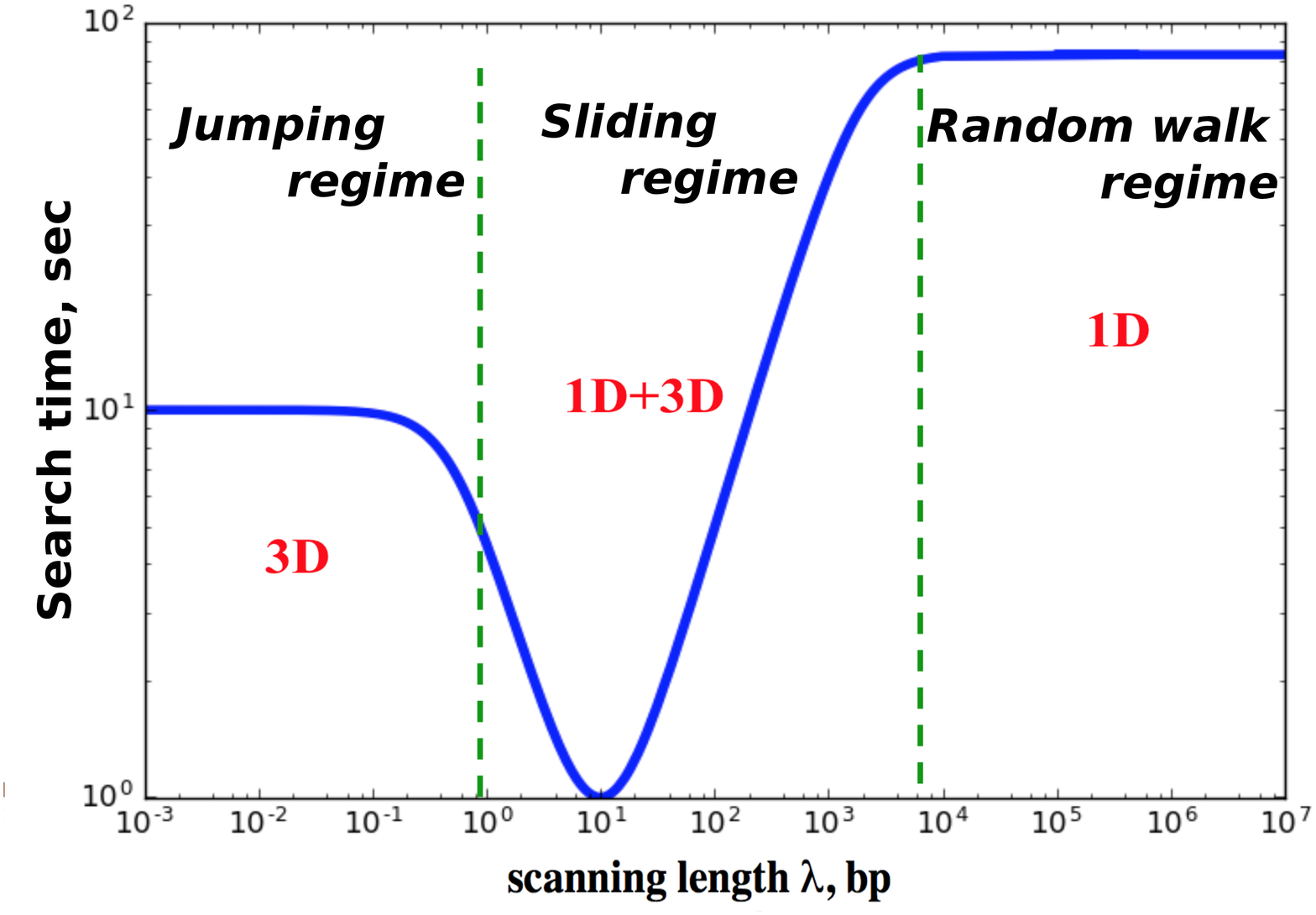}}
\caption{Mean search times as a function of the scanning length parameter $\lambda = \sqrt{u/k_{off}}$. The parameters utilized in calculations are: $L = 10^3$ bp, $u = k_{on} =10^5$ $s^{-1}$, and $m = L/2$. The transition rate $k_{off}$ is varied to change $\lambda$.}
\label{fig::phase_diag}
\end{figure}

The results of our calculations for the mean search times are presented in Figure \ref{fig::phase_diag}. Our main finding here is that there are three dynamic search regimes depending on the values of kinetic parameters. It is convenient to introduce here a scanning length $\lambda=\sqrt{u/k_{off}}$, which gives the average distance that the protein molecule travels on DNA during each search cycle. This quantity is related to the parameter $S_{1}(0)$, but it is not the same because the protein might visit the same sites several times. If the protein molecule has a strong affinity to bind non-specifically to the DNA molecule (small $k_{off}$, $\lambda > L$), then there will be only one searching cycle. After binding to DNA the protein will not dissociate until it finds the target. In this case, the mean search time scales as $\sim L^{2}$ because the DNA-bound protein does a simple unbiased random walk. We call this dynamic phase a random-walk regime. Because of the redundancy of the random walk the search in this regime should be generally slow: many sites are repeatedly visited. In the opposite limit of weak attractions between DNA and protein molecules (large $k_{off}$, $ \lambda <1$), the protein can bind to DNA but it cannot slide because it quickly dissociates back into the solution. The protein on average makes $L$ searching cycles ($T_{0} \sim L$). This dynamic regime is called a jumping regime.  The search in this regime is generally fast as long as the associations are also fast. The most interesting behavior is observed for the intermediate interactions, which we label as a sliding regime. Here the scanning length $\lambda$ is larger than one but smaller than the length of DNA $L$, and the number of searching cycles is also proportional to $L$. But in this regime the system can reach the most optimal dynamic behavior with the smallest search times. This search facilitation is achieved due to the fact that the fluxes to the target are coming now from both the bulk solution and from the DNA chain. This is one of the main mechanisms of the facilitated diffusion of proteins during the target search, but other processes like inter-segment transfer might also contribute significantly in the facilitated diffusion \cite{esadze14}.

\section{The Effect of Multiple Targets and Traps}

The advantage of the discrete-state stochastic framework with the first-passage analysis presented above  is that it can be extended and generalized to more realistic biological situations. This allows us to investigate  important questions related to the mechanisms of the protein target search on DNA. Let us present several specific examples, although many more results have been obtained.\cite{veksler13,shvets15,shvets16,shvets15b,shvets16b,shvets17,esadze14,kochugaeva16,kochugaeva17,lange15a,lange15b,shin18,kochugaeva17a}  We start with the problem of how the presence of multiple target sites or multiple semi-specific trap sites affect the dynamics of the protein search.

\begin{figure}
\centerline{\includegraphics[width=9cm]{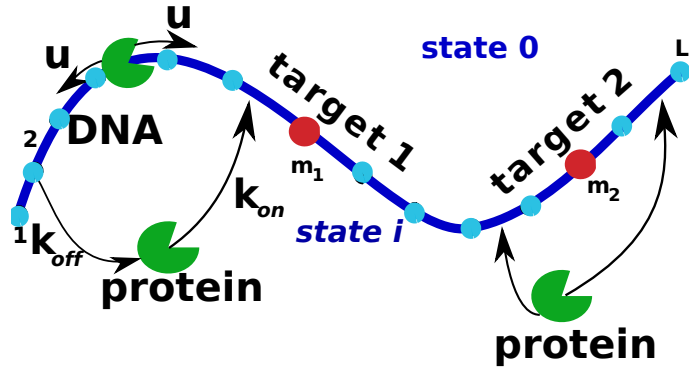}}
\caption{A schematic view of the discrete-state stochastic  model of the protein search with multiple specific sites. Targets are located at the sites $m_{1}$ and $m_{2}$. }
\label{Fig3}
\end{figure}

It is known that in eukaryotic cells multiple target sites are available on the accessible DNA fragments \cite{lodish,alberts,phillips,townson07}. The protein search is accomplished in these systems  when the protein molecules finds for the first time {\it any} of the target sites. It has been argued that the mean search time in this system might not decrease proportionally to the number of targets as one would naively expect from simple-minded applications of chemical kinetics \cite{lange15a}. This is due to the complex mechanism of the protein search that involves both 3D and 1D motions \cite{lange15a}. Applying our discrete-state stochastic framework to this problem, we consider a model with multiple targets at arbitrary locations as presented in Figure \ref{Fig3}. To describe the search dynamics in this system, we again introduce the first-passage probability function $F_{n}(t)$ of finding {\it any} of the targets at time $t$ if the process started at $t=0$ at the site $n$. Targets are dividing the DNA chain into several homogeneous segments, and this allows us to solve the corresponding backward master equations  as explained in Section 2. This leads to the following explicit expression for the mean search time for any number of targets \cite{lange15a},
\begin{equation}
    T_{0}=\frac{1}{k_{on}}\frac{L}{S_{i}(0)}+\frac{1}{k_{off}}\frac{L-S_{i}(0)}{S{i}(0)},
\end{equation}
with a function $S_{i}(0)$ describing the average number of distinct sites scanned by the protein on DNA with $i$ targets. This formula is a generalization of Equation (\ref{eq13}) when there is only one target  ($i=1$). Specific expressions for $S_{i}(0)$ for various numbers of randomly distributed targets have been obtained \cite{lange15a}. For example, for $i=2$ it was shown that
\begin{equation}\label{eqS2}
    S_{2}(s)=\frac{(1+y)\left[2(1-y^{2L+m_{1}-m_{2}})+(1-y^{m_{2}-m_{1}})(y^{2m_{1}-1}+y^{1+2(L-m_{2})})\right]}{(1-y)(1+y^{2m_{1}-1})(1+y^{1+2(L-m_{2})})(1+y^{m_{2}-m_{1}})},
\end{equation}
where the parameter $y$ is given in Equation \ref{eq_y}.

\begin{figure}[htbp]
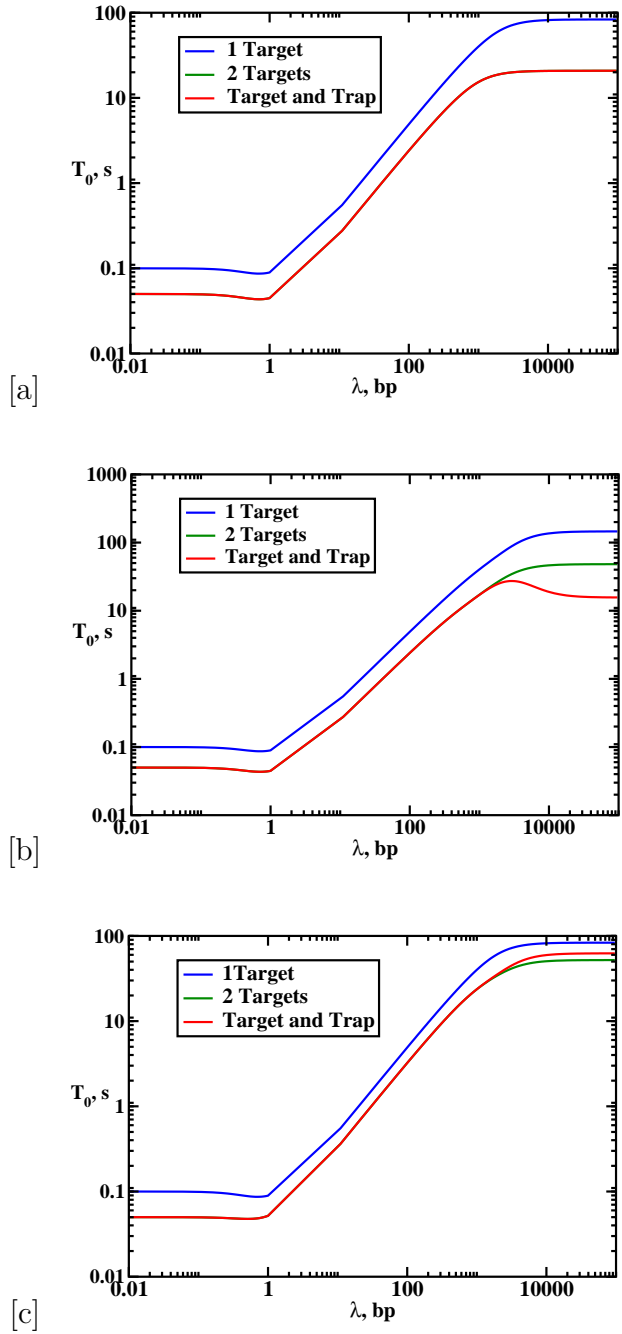

	\centering
	[a] \hskip 0.1 in \includegraphics[scale=0.30]{figures/Fig3-3a.eps} \vskip 0.3in
        [b] \hskip 0.1 in \includegraphics[scale=0.30]{figures/Fig3-3b.eps} \vskip 0.3in
        [c] \hskip 0.1 in \includegraphics[scale=0.30]{figures/Fig3-3c.eps} \vskip 0.3in
	\caption{Dynamic phase diagrams for the protein search on DNA with one target at the position $m$, with two targets at the positions $m_{1}$ and $m_{2}$ and with the target and the trap  at the positions $m_{1}$ and $m_{2}$, respectively. Parameters used for calculations are: $k_{on}=u=10^5$  s$^{-1}$ and $L=10000$. a) $m=L/2$, $m_{1}=L/4$ and $m_{2}=3L/4$; b) $m=L/4$, $m_{1}=L/4$ and $m_{2}=L/2$; and c) $m=L/2$, $m_{1}=L/2$ and $m_{2}=L$. Adapted with permission from Ref. \cite{lange15b}.}
    \label{fig4}
\end{figure}

\begin{figure}
\centerline{\includegraphics[width=8.5cm]{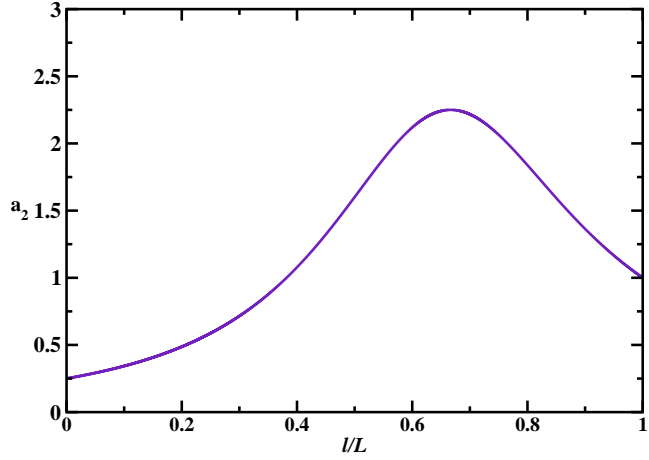}}
\caption{Ratio of the mean search times as a function of the normalized distance between the targets for single-target and two-target systems ($l$ is the distance between between targets, $L$ is the DNA length). The single target is in the middle of the chain. In the two-target system, one of the specific sites is fixed at the end and the position of the second one is varied. The parameters used in calculations are: $u=k_{on}=10^{6}$ s$^{-1}$; $k_{off}=10^{-4}$ s$^{-1}$; and $L=10000$. Adapted with permission from Ref. \cite{lange15a}.}
\label{Fig5}
\end{figure}

To understand the effect of multiple targets on the protein search dynamics, we analyze the results of explicit calculations for mean search times as presented in Figure \ref{fig4}. It is found that the presence of multiple targets does not affect the overall dynamic phase diagram as compared with the single-target case: three search regimes are again observed depending on the size of the scanning length, the target size and the size of the DNA segment. Generally, the search is faster in the multiple-target systems. However, surprisingly, increasing the number of specific sites  might not always accelerate the search. To quantify this effect, we introduced an acceleration parameter, $a_{n}=T_{0}(1)/T_{0}(n)$, where $T_{0}(n)$ is the mean search for the system with $n$ targets. This ratio gives a numerical value of how faster the search is in the presence of $n$ targets in comparison with the single-target system. It is illustrated in Figure \ref{Fig5}. One can see that there is a range of parameters when the search dynamics in the system with two targets can be slower than the dynamics in the system with one target. This happens in the effectively 1D search regime (random-walk dynamic phase) when the single target is located in the middle of the DNA chain, while two targets are close to each other and located near one of the ends of the DNA segment.  In this case, for the protein molecule the two targets are viewed as effectively a single target site (with the size equal to two target sites) because they are so close to each other. But it is faster to find the target located in the middle of the chain than the target positioned near the ends.\cite{veksler13} This is the main reason why having multiple targets does not always lead to decrease in the search times. Thus, our theoretical analysis predicts that the degree of acceleration due to the presence of multiple targets depends on the nature of the dynamic search phase and on the location of the specific sites with respect to each other and with respect to the middle point of DNA \cite{lange15a}.

Another important factor that might affect the protein search dynamics is the existence of so-called semi-specific sites, or decoys, on DNA. These sites have a chemical composition very similar to the specific targets with differences in only one or few nucleotides. The protein molecule can be trapped in these sites, and this should influence the search for real targets. To analyze this effect, we  can extend the simplest model to include the possibility of traps, assuming that associations to these semi-specific sites are effectively irreversible \cite{lange15b}. This assumption is reasonable because the search times in many systems are  relatively short and the experimental observations also limited in time. Thus the bindings to decoys can be viewed as effectively irreversible. The first-passage analysis can be applied here, but we have to notice that only a fraction of trajectories will reach the correct target site. Then the main quantity of our calculations, the first-passage probability function $F_{n}(t)$, is now a {\it conditional} probability for the protein molecules  not captured by the trap to find the target site. 

Let us consider a system consisting of a single target at the site $m_{1}$ and a single trap at the site $m_{2}$ on the DNA molecule with $L$ sites \cite{lange15b}. The scheme presented in Figure {\ref{Fig3}} is also a correct representation of this system with the correction that instead of the second target there is a trap in the site $m_{2}$, and the successful search corresponds to the protein molecule finding the specific site $m_{1}$. Following our theoretical method, the corresponding backward master equations can be solved and they yield the Laplace transform of the first-passage probability function to find the target if the protein starts from the bulk solution \cite{lange15b},
\begin{equation}
\widetilde{F_{0}(s)}=\frac{k_{on}(k_{off}+s)S_{0}(s)}{L s(k_{off}+k_{on}+s)+k_{off}k_{on}S_{2}(s)}, \end{equation}
with 
\begin{equation}
    S_{0}(s)=\frac{(1+y)(1-y^{m_{1}+m_{2}-1})}{(1-y)(1+y^{2m_{1}-1})(1+y^{m_{1}-m_{2}})},
\end{equation}
and the parameters $y$ and $S_{2}$ given in Equations (\ref{eq_y}) and (\ref{eqS2}), respectively. This allows us to evaluate all dynamic properties in the system and to test the effect of traps. 

The probability to reach the target (i.e., the fraction of the successful trajectories) is now given by a so-called splitting probability function \cite{vankampen,redner},
\begin{equation}
    \Pi \equiv \widetilde{F_{0}(s=0)}=\frac{S_{0}(0)}{S_{2}(0)}.
\end{equation}
The mean search time, which is the conditional mean first-passage time to reach the target, can be estimated by averaging over the successful trajectories, producing
\begin{equation}
T_0 \equiv -\frac{\frac{\partial \widetilde{F_0(s)}}{\partial s}\bigg\vert_{s=0}}{\Pi}=\frac{1}{k_{on}}\frac{L}{S_{2}(0)}+\frac{1}{k_{off}}\frac{L-S_{2}(0)}{S_{2}(0)} +\Pi \frac{\partial}{\partial s} \left[ \frac{S_{2}(s)}{S_{0}(s)} \right] \bigg\vert_{s=0}.
\end{equation}
Let us analyze this expression. On the left side, the division by the splitting probability emphasizes the fact that this is the conditional mean search time. It is also  interesting to note that the first two terms on the right side of the equation is exactly the mean search time for the system with two targets  and no traps (at the sites $m_{1}$ and $m_{2}$) as we discussed above \cite{lange15a}, while the third term is a correction which accounts for the fact that the site at $m_{2}$ is actually the trap. The main reason for this is the observation that the sites $m_{1}$ and $m_{2}$ are special locations where all trajectories are end up in both systems, with two targets and with the target and the trap. For the two-target case the mean search times are averaged over all trajectories to both sites, while for the target and the trap system the mean search times are obtained only by considering the trajectories finishing at the target \cite{lange15b}.

\begin{figure} 
	\centerline{
	\includegraphics[width=8.5cm]{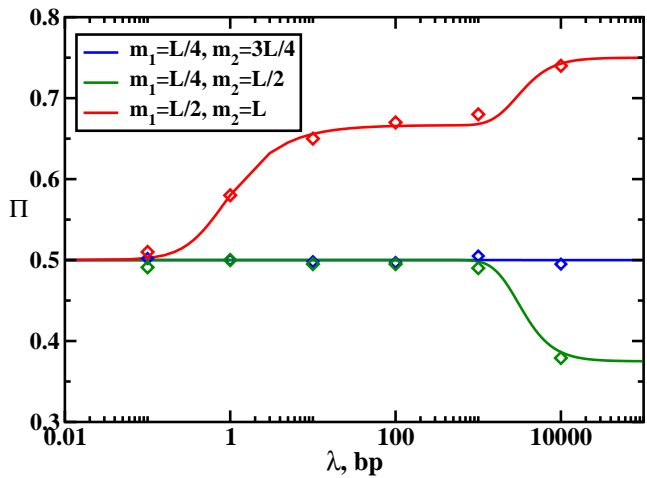}} 
        \caption{Probability to reach the target as a function of the scanning length for different distributions of the target and trap sites. Parameters used for calculations are: $k_{on}=u=10^5$  s$^{-1}$, $L=10000$ and $k_{off}$ is changing. Symbols are from Monte Carlo computer simulations. Adapted with permission from Ref. \cite{lange15b}}\label{fig6}
\end{figure}

The results of calculations for the dynamic properties of the protein search in the presence of traps are presented in Figures \ref{fig4} and \ref{fig6}. Again, three dynamic search phases are observed, but adding the trap generally facilitates the search dynamics, which is a counter-intuitive result: see Figure \ref{fig4}. However, this acceleration (in comparison with the single-target system) is always associated with lowering of the probability of reaching the specific target, as shown in Figure \ref{fig6}. This means that the protein molecules might reach the target faster in the presence of the traps, but the fraction of such events is decreasing. In addition, the search dynamics is sensitive to the nature of the dynamic phase. The strongest effect due to the presence of the trap is observed in the effective 1D random-walk regime (because it has only one searching cycle) where the locations of the target and the trap strongly influence the search. In other dynamic regimes, the effect is smaller.

\section{Sequence heterogeneity}

Real DNA molecules  are heterogeneous polymers consisting of several types of subunits. This means that the interactions between protein and DNA molecules depend on the DNA sequence at the location where they meet. It is reasonable to expect that this sequence dependence in the interaction strength  should affect the protein search dynamics because the diffusion rate for the non-specifically bound proteins will be position-dependent \cite{phillips,mirny09,bauer15}. Similarly, association and dissociation rates should also depend on the location of the protein molecule on DNA. In addition, recent theoretical investigations suggested that different DNA sequence symmetries might lead to additional effective interactions between protein and DNA molecules \cite{afek11,afek12,afek13,afek14}. The discrete-state stochastic framework with the first-passage analysis is a convenient tool to investigate the effect of DNA sequence heterogeneity and symmetry on the protein search dynamics \cite{shvets15}.

\begin{figure} 
	\centerline{
	\includegraphics[width=8.5cm]{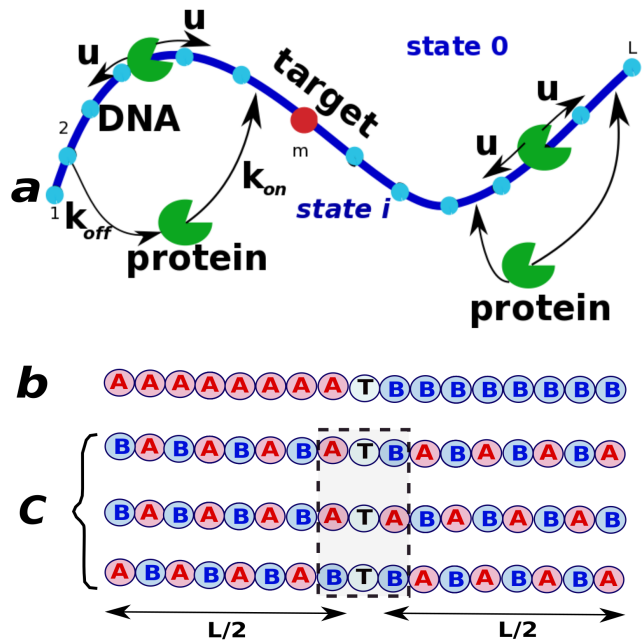}} 
        \caption{A simplified view on the protein search on DNA with two different types of subunits, $A$ and $B$. a) A general scheme; b) DNA is viewed as a symmetric block copolymer with the target in the middle of the chain; c) DNA is viewed as alternating copolymer with different compositions of the subunits flanking the target in the middle of the chain. Adapted with permission from Ref. \cite{shvets15}}\label{fig7}
\end{figure}

Our goal here is clarify the molecular origin of how the sequence heterogeneity influences the protein target search.  We assume here a simplified picture of DNA, in which each monomer  can be one of two chemical species, $A$ or $B$, as presented in Figure {\ref{fig7}} \cite{shvets15}. When the protein is bound to the subunit $A$ ($B$), it interacts with energy $\varepsilon_A$ ($\varepsilon_B$), and the difference between interaction energies is given by a parameter  $\varepsilon = \varepsilon_A - \varepsilon_B \ge 0$. This means that the protein attracts stronger to the $B$ sites than to the $A$ sites. The protein molecule can diffuse along DNA with a rate $u_A \equiv u$ or $u_B=ue^{-\varepsilon}$, where $\varepsilon$ is measured in $k_{B}T$ units. This reflects the assumption that if the protein interacts stronger with the DNA at given location then it will move out of this site slower. In addition, we assume that, independently of the chemical nature of the neighboring sites, sliding out of the sites $A$ is characterized by the rate $u_A$, while the diffusion out of the sites $B$ is given by $u_B$. From the bulk solution the protein might associate to any site $A$ or $B$ on DNA with the corresponding rates $k^A_{on}=k_{on}$ or $k^B_{on}=k_{on}e^{-\theta\varepsilon}$. Note that for convenience the on-rates defined here as the rates per unit site, in contrast to our definitions in the previous sections. Similarly, the dissociations from the DNA chain are described by the rates $k^A_{off}=k_{off}$ and $k^B_{off}=k_{off}e^{(\theta-1)\varepsilon}$. Here, the parameter $0\le \theta\le1 $ specifies how the protein-DNA interaction energy is distributed between the association and dissociation transitions \cite{shvets15}. The physical meaning of this parameter is that the protein molecule tends to bind faster and to dissociate slower from the stronger attracting sites $B$, as compared with the weaker attracting $A$ sites. The parameter $\theta$ accounts for these effects.

To quantify the role of  sequence heterogeneity, we consider the DNA molecule with a fixed chemical composition (the fractions of $A$ and $B$ monomers are the same), but with different arrangements of subunits. Two limiting cases are specifically analyzed. One of them views the DNA molecule as two homogeneous segments of only $A$ and only $B$ subunits separated by the target in the middle of the chain (Figure \ref{fig7}). Another one is the DNA chain with the alternating $A$ and $B$ sites. The block copolymer has two homogeneous sequence segments, while the alternating polymers are more heterogeneous. It is important to note that in both cases, the overall interaction between the protein and DNA is the same (because the overall chemical composition in both cases is identical), and thus our analysis probes only the effect of the heterogeneity and symmetry in the subunit positions, in contrast to other theoretical treatments \cite{brackley12}.

\begin{figure}
\centerline{\includegraphics[width=8.5cm]{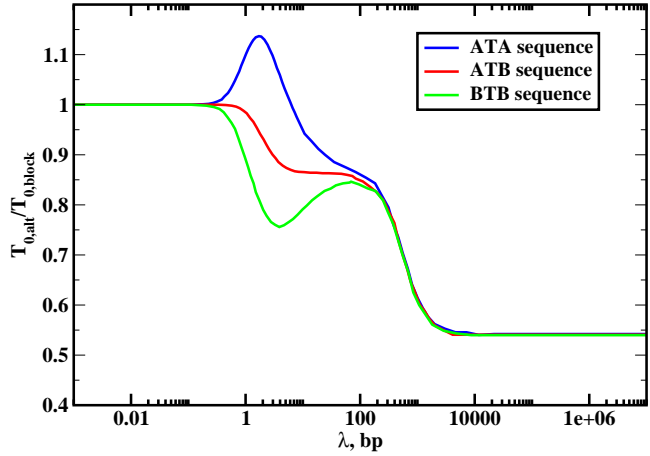}}
\caption{The ratio of the mean search times for the alternating DNA sequences and for the block copolymer DNA sequences as a function of the scanning length $\lambda=\sqrt{u/k_{off}}$. Three different chemical compositions near the target ($T$) are distinguished, namely, $ATA$, $ATB$, $BTB$. The transition rates are $u = 10^5\, s^{-1}$ and $k_{on} = 0.1\, s^{-1}$. The DNA length is $L = 1000$, the loading parameter is $\theta = 0.5$, and the energy difference of interactions for the protein with $A$ and $B$ sites is $\varepsilon=5$ $k_{B}T$. Adapted with permission from Ref. \cite{shvets15}.}
\label{fig::seq_rel_time}
\end{figure}

Applying again the first-passage approach and solving the corresponding equations leads to the explicit expressions for mean search times for all situations shown in Figure \ref{fig7} \cite{shvets15}. For example, for the block copolymer DNA sequences, we obtain
\begin{equation}
T_{0}=\frac{k_{off}+k_{on}\left[(L/2-P_{A})+e^{\varepsilon}(L/2-P_{B})\right]}{k_{on} k_{off}(1+P_{A}+e^{\theta \varepsilon} P_{B})},
\end{equation}
where
\begin{equation}
    P_{i}=\frac{x_{i}^{1-L/2}-x_{i}^{1+L/2}}{(1-x_{i})(x_{i}^{1+L/2}+x_{i}^{_L/2})},
\end{equation}
\begin{equation}
x_{i}=\frac{2u_{i}+k_{off}^{(i)}-\sqrt{(2u_{i}+k_{off}^{(i)})^{2}-4u_{i}^{2}}}{2u_{i}},
\end{equation}
for $i=A$ or $B$. The expressions for the mean search time for alternating sequences are quite bulky and can be found in Ref. \cite{shvets15}.

The results of our calculations are presented in Figure \ref{fig::seq_rel_time}, where the ratio of the mean search times for the block copolymer and alternating sequences are plotted. The analysis of this figure produces several interesting observations. First, we see that three dynamic search regimes are also found in this system and the effect of sequence heterogeneity on protein search dynamics depends on the nature of the dynamic phase. In the jumping regime when the protein does not slide along the DNA contour ($\lambda <1$), the symmetry of the sequence does not play any role. This is because in this case the process is taking place only via associations and dissociations (3D search), and the structure of the DNA chain is not important. The situation is different for the intermediate sliding regime (3D+1D search, $1 < \lambda < L$) where in most cases, the search on alternating sequences is faster. This can be explained by noticing that the search time in this dynamic phase is proportional to $L/\lambda$, which gives the average number of cycles before the protein can find the target. In the block copolymer sequence, the protein mostly comes to the target from the $B$ segment because of stronger interactions with these sites, i.e, it comes from one side of the DNA molecule. In the alternating sequences, the protein can reach the target from both sides of DNA, and this lowers the overall search time. It can be shown analytically that the scanning length on the alternating segment is larger than the scanning length for the $B$ segment, i.e., $\lambda_{AB} > \lambda_B$ \cite{shvets15}. Then the search is faster for the alternating sequences because $L/\lambda_{AB} < L/\lambda_B$, i.e., the number of searching cycles is lower for the alternating sequences, which helps to find the target faster. The only deviation from this picture is found for $ATA$ sequences, which corresponds to having two $A$ sites around the target site, where for the small range of parameters the search is slower than in the block copolymer sequence. This effect can be explained by the fact that the protein does not sit at $A$ sites for the long time and it moves quickly away, effectively increasing the barrier to enter the target via DNA \cite{shvets15}. Thus, our theory predicts that the composition of the DNA flanking sites around the target sequences might affect the dynamics of reaching them. It is interesting to note that recent experiments are consistent with our theoretical predictions \cite{le18}.

In the random-walk regime (1D search, $\lambda > L$), the effect of the sequence heterogeneity is even stronger. The protein molecule finds the specific binding site up to 2 times faster for more heterogeneous alternating DNA sequences. To understand this behavior, we note that in this case the mean first-passage time to reach the target is a sum of residence times on the DNA sites since the protein will not dissociate until the target is located so that all trajectories to the target are one-dimensional. Because the target is in the middle of the chain, the mean time to reach the target from the block copolymer sequence can be approximated as $T_0 \simeq (L/4) \tau_B$, where $\tau_B$ is the average residence time on any site $B$. The protein prefers to start the search at any position on the $B$ segment with equal probability, i.e., the distance to the target varies from 0 to $L/2$. Then, the average starting position of the protein is $L/4$ sites away from the target. For the alternating sequences, the average distance to the target is approximately the same ($L/4$), but the chemical composition of intermediate sites on the path to the target is different, yielding, $T_0 \simeq (L/8) \tau_A + (L/8) \tau_B$ ($\tau_{A}$ is the residence time on $A$ sites). The protein spends much less time on $A$ subunits, and this leads to faster search for the alternating DNA sequences. For $\tau_A \ll \tau_B$, this also explains the factor of 2 in the search speed. In this case, the $B$ subunits can be viewed as effective traps that slow down the search dynamics. Thus, our theoretical calculations make surprising predictions that the sequence heterogeneity almost always lead to faster protein search for targets on DNA despite the fact that it lowers the effective protein-DNA binding affinity \cite{afek11,afek12,afek13,afek14}. And the stronger the contribution of the 1D search modes, the more relevant will be the effect of sequence heterogeneity.

\section{The Effect of Crowding on DNA in the Protein Target Search}

Living cells are typically crowded with a large number of molecules, and many of them are attached to the DNA chains \cite{alberts,lodish}. This should prevent the fast protein search for targets on DNA, and earlier theoretical studies supported this prediction \cite{gomez16}. However, surprisingly, experiments show that crowding on DNA does not affect much the effectiveness of the protein target search \cite{hammar12,mahmutovic15}, and this was also found in MD simulations \cite{marcovitz13}. By applying the discrete-state stochastic approach, we were able to clarify the role of the crowding on DNA in the protein target search.

\begin{figure}
\centerline{\includegraphics[width=8.5cm]{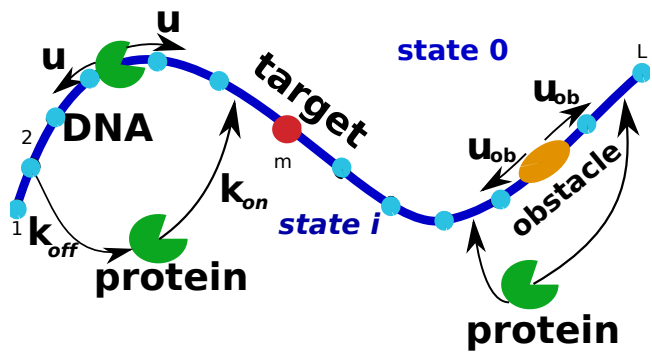}}
\caption{A schematic view of the protein target search in the presence of a moving obstacle on DNA. The crowding particle cannot dissociate from DNA, while the protein molecule can dissociate into the solution, labeled as state $0$, and return back to the DNA chain.} \label{fig9}
\end{figure}

To analyze this problem, the model illustrated in Figure {\ref{fig9}}  is considered. There is a single DNA molecule with $L+1$ binding sites, and one of them is the target (at the site $m$). On the DNA chain there is also a crowding particle that can diffuse with a rate $u_{ob}$, but it cannot leave DNA. A single protein molecule starts from the solution (state 0) and it can bind to any site on DNA that is not occupied by the crowder with a rate $k_{on}$ (rate per site). The bound protein molecule can diffuse with a rate $u$, and there is an exclusion interaction between the protein and the crowder. Finally, the protein molecule can dissociate from DNA to the bulk solution with a rate $k_{off}$: see Figure {\ref{fig9}}.

\begin{figure}
\centerline{\includegraphics[width=8.5cm]{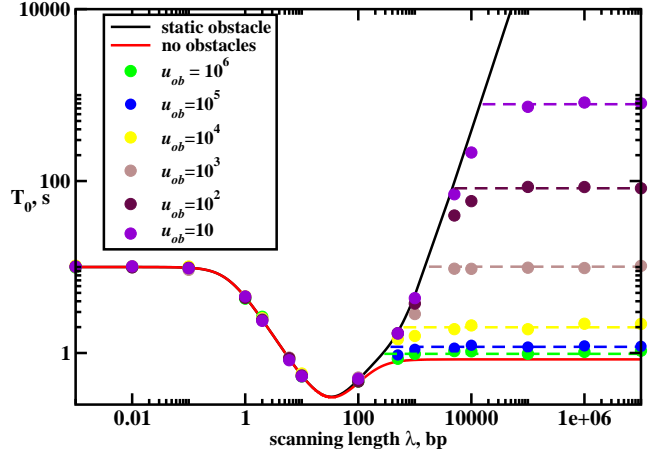}}
\caption{ Mean search times to find the target in the system with a mobile crowder on DNA. The DNA chain has $L=1000$ sites, and the target is in the middle of the chain, $m=L/2$. Parameters used for calculations are $k_{on}=0.1$ s$^{-1}$, $u=10^{5}$ s$^{-1}$ and variable $u_{ob}$. Solid curves correspond to analytical results for DNA without obstacles and for DNA with a static obstacle, which are averaged over all initial positions of the crowder. Symbols correspond to Monte Carlo computer simulations. Dashed lines describe the approximate theory, as explained in the text. Adapted with permission from Ref. \cite{shvets16}.}\label{fig10}
\end{figure}

Investigating the model with the mobile crowding particle on DNA first using Monte Carlo computer simulations, it is found that there are three search regimes depending on the main length scales  in the system. This is shown in Figure {\ref{fig10}} for the mean search times to find the target as a function the scanning length $\lambda$.  We can understand the complex dynamics in this system using the following arguments. If the diffusion rate of the crowder is much smaller than other rates  ($u_{ob} \ll u$, $k_{on}$ and $k_{off}$), then the protein molecule will find the target before the crowding particle can move away from its original location. But we already explicitly solved the problem of the protein target search with static obstacles using the same discrete-state stochastic approach with the first-passage analysis \cite{shvets16b}. Then the mean search time in the system with movable crowder can be approximated as the average over all possible static locations of the crowding particle \cite{shvets16}, yielding
\begin{equation}
\langle T_{0} \rangle \simeq \frac{1}{L} \left( \sum_{l_{ob}=1}^{m-1} T_{ob}(l_{ob})+\sum_{l_{ob}=1}^{L-m}T_{ob}(l_{ob}) \right),
\end{equation}
where
\begin{equation}
T_{ob}=\frac{k_{off}+k_{on}(L-S_{ob}(0))}{k_{on}k_{off}S_{ob}(0)},
\end{equation}
is the mean search time with the static obstacle located at a distance $l_{ob}$ from the target. An auxiliary function $S_{ob}$ is given by \cite{shvets16b}
\begin{equation}
S_{ob}(s)=\frac{y(y^{-m}-y^{m})}{(1-y)(y^{m}+y^{1+m})}+\frac{y(1-y^{2l_{ob}-2})}{(1-y)(1+y^{2l_{ob}-1})}
\end{equation}
with the parameter $y$ specified in Equation {\ref{eq_y}}.

This simple approximate theory works quite well in the dynamic regimes where 3D pathways are important for the search ($ \lambda < L$).  However, theoretical arguments fail in the random-walk regime where 1D dynamics dominate the search. These results are expected. The protein molecule that collides with the crowding particle on DNA in dynamic regimes with 3D pathways will have the opportunity to dissociate into the bulk solution and to avoid the blocking effect.  But in the random-walk regime (1D search) there is no such opportunity, and the search times will definitely increase. Computer simulations also indicate that the search times in this regime depend on the diffusivity of the crowding particle. The search is faster for more mobile crowders: see Figure {\ref{fig10}}.

The dynamics in the random-walk regime can be explained using the following arguments. The overall search can be viewed as consisting of two terms,
\begin{equation}
\langle T_{0}^{ob} \rangle \simeq T_{0}+ \langle T_{bl} \rangle,
\end{equation}
where $T_{0}$ is the search in the random-walk regime without any crowders, and it is given in Equation {\ref{eq13}}. The second term is the average time it takes for the crowder to diffuse away and clear the path for the protein to reach the target without interference \cite{shvets16}. It was shown that this blocking time $T_{bl}$ depends on the location of the target and the diffusion rate of the crowding particle $u_{ob}$ \cite{shvets16},
\begin{equation}
\langle T_{bl} \rangle = \frac{m^{4}+(L-m)^{4}}{16 u_{ob}(L^{2}+m^{2}-mL)}.
\end{equation}

This simple theoretical arguments show excellent agreement with Monte Carlo computer simulations: see dashed lines in Figure {\ref{fig10}}. But more importantly, they provide a clear molecular picture on the role of the crowding  on DNA in the protein target search.  If the protein search is dominated by 1D pathways and the mobility of the crowder is low the search dynamics will be significantly slowed down. But if the search involves mostly 3D pathways and the crowder is mobile the mean search times will not be affected much. It seems that real biological systems operate in 3D+1D regime, and crowding particles diffuse with the rates comparable to the searching proteins ($u \sim u_{ob}$) \cite{phillips}. Then one might conclude that the effect of the crowders on DNA should be minimal. This fully agrees with experimental observations and with results from MD simulations \cite{mahmutovic15,marcovitz13}.

\section{Conclusions and Future Directions}

Although protein search for targets on DNA is a very complex phenomenon that involves multiple biochemical and biophysical processes, significant advances in our understanding of the underlying molecular mechanisms have been achieved in recent years. A major role in this success is due to analysis of the systems using the discrete-state stochastic framework supplemented by explicit calculations via the first-passage probabilities method. In this review, we presented and explained this theoretical approach by considering the protein target search in various systems. It is important to emphasize that the main advantage of our theoretical approach is the ability to obtain analytical results that clarify the physics of the underlying processes. In addition, the method can be easily extended in many directions, as shown in this work, as well as in other cases which we did not discuss in this work, such as  the role of conformational transitions \cite{kochugaeva16} and the effect of DNA loop formation in the protein target search \cite{shvets16b}. Furthermore, our theoretical calculations using this theoretical framework were successful in explaining the experimental observations on homology search by RecA protein filaments \cite{kochugaeva17a} and the dynamics of CRISPR genome interrogation \cite{shvets17}.

Several important dynamic features of the protein search for targets on DNA have been identified from theoretical analysis. It is found that the dynamic phase diagram of the protein target search always shows thee dynamic regimes, which are determined by the three relevant length scales in the system: the size of DNA, the average scanning length of the non-specifically bound proteins, and the size of the target sequence. Depending on the dynamic phase, the search is dominated by the 3D motions (jumping regime), 1D motions (random-walk regime) or  a combination of 3D and 1D motions in the sliding regime.  The analysis shows that the most optimal search dynamics can be achieved in the dynamic regime when the protein molecules explore both 1D and 3D pathways during the search. In this case, the protein can reach the target by sliding from the DNA chain or by directly binding from the solutions. Theoretical calculations also indicate that the presence of several target sites influences the search dynamics differently depending on the locations of the targets on DNA and distances between them.  Surprising observations are found in the system with semi-specific sites, which are viewed as effective traps. It is shown that the search dynamics can be faster in this case, but it comes with the price of lowering the yield of the protein molecules reaching the target. We also investigated the effect of sequence heterogeneity and symmetry in the protein search dynamics. Our calculations indicate that the search is faster for more heterogeneous sequences, and the chemical composition around the target is also an important factor in this process.  Furthermore, our method allowed us to probe the effect of crowding on DNA in the protein target search. It is shown that it depends on the dynamic phase and on the mobility of the crowding particles. The crowders influence the protein search stronger when 1D pathways dominate and when the diffusivity of the crowding particle is small enough so that the protein will be frequently  blocked during the process. Increasing the mobility of the crowders and/or increasing the contribution of 3D search pathways lowers the effect of the crowding. These theoretical arguments fully agree with experimental observations and MD computer simulations.

Despite tremendous progress in theoretical understanding of the protein target search phenomena, there are many questions remain on the molecular mechanisms of these processes. It is still unclear what is the nature of protein-DNA interactions in the regions surrounding the target sequences. Is the effective potential created by these interactions drives the protein molecule to the target like a funnel or is it completely random? How large is the size of the flanking segments that affect the finding of the target? What is the role of DNA topology in the protein target search? This is especially important for proteins that have several binding sites for DNA which can form DNA loops and other complex structures. Another interesting question is the role of various DNA and protein conformations in these processes.  It is clear that further progress in understanding protein target search phenomena depends on combining theoretical, computational and experimental methods.

\acknowledgement
A.B.K. acknowledges the support from the Welch Foundation (C-1559), from the NSF (CHE-1664218) and from the Center for Theoretical Biological Physics sponsored by the NSF (PHY-1427654).

\end{document}